\documentclass[aps,prl,twocolumn,showpacs,preprintnumbers,amsmath,amssymb,superscriptaddress]{revtex4}%

\usepackage{graphicx}
\usepackage{dcolumn}
\usepackage{bm}
\usepackage{color}

\begin{document}

\preprint{preprint(\today)}


\title{Pressure tuning of structure, superconductivity and novel magnetic order in the Ce-underdoped electron-doped cuprate T$^{'}$-Pr$_{1.3-x}$La$_{0.7}$Ce$_{x}$CuO$_{4}$ ($x$ = 0.1)}

\author{Z.~Guguchia}
\email{zg2268@columbia.edu;zurab.guguchia@psi.ch} 
\affiliation{Department of Physics, Columbia University, New York, NY 10027, USA}
\affiliation{Laboratory for Muon Spin Spectroscopy, Paul Scherrer Institute, CH-5232
Villigen PSI, Switzerland}

\author{T.~Adachi}
\affiliation{Department of Engineering and Applied Sciences, Sophia University, 7-1 Kioi-cho, Chiyoda-ku, Tokyo 102-8554, Japan}

\author{Z.~Shermadini}
\affiliation{Laboratory for Muon Spin Spectroscopy, Paul Scherrer Institute, CH-5232
Villigen PSI, Switzerland}

\author{T.~Ohgi}
\affiliation{Department of Applied Physics, Tohoku University, 6-6-05 Aoba, Aramaki, Aoba-ku, Sendai 980-8579, Japan}

\author{J.~Chang}
\affiliation{Physik-Institut der Universit\"{a}t Z\"{u}rich, Winterthurerstrasse 190, CH-8057 Z\"{u}rich, Switzerland}

\author{E. Bozin}
\affiliation{Condensed Matter Physics and Materials Science Department,
Brookhaven National Laboratory, Upton, NY 11973, USA}

\author{F.~von Rohr}
\affiliation{Physik-Institut der Universit\"{a}t Z\"{u}rich, Winterthurerstrasse 190, CH-8057 Z\"{u}rich, Switzerland}

\author{A.M. dos Santos}
\affiliation{Neutron Sciences Directorate, Oak Ridge National Laboratory, Oak Ridge, Tennessee 37831, USA}

\author{J.J. Molaison}
\affiliation{Neutron Sciences Directorate, Oak Ridge National Laboratory, Oak Ridge, Tennessee 37831, USA}

\author{R. Boehler}
\affiliation{Neutron Sciences Directorate, Oak Ridge National Laboratory, Oak Ridge, Tennessee 37831,
USA}
\affiliation{Carnegie Institution of Washington, USA}

\author{Y.~Koike}
\affiliation{Department of Applied Physics, Tohoku University, 6-6-05 Aoba, Aramaki, Aoba-ku, Sendai 980-8579, Japan}

\author{A.R.~Wieteska}
\affiliation{Department of Physics, Columbia University, New York, NY 10027, USA}

\author{B.A. Frandsen}
\affiliation{Department of Physics, Columbia University, New York, NY 10027, USA}

\author{E.~Morenzoni}
\affiliation{Laboratory for Muon Spin Spectroscopy, Paul Scherrer Institute, CH-5232
Villigen PSI, Switzerland}

\author{A.~Amato}
\affiliation{Laboratory for Muon Spin Spectroscopy, Paul Scherrer Institute, CH-5232
Villigen PSI, Switzerland}

\author{S.J.L.~Billinge}
\affiliation{Condensed Matter Physics and Materials Science Department, Brookhaven National Laboratory, Upton, NY 11973, USA}
\affiliation{Department of Applied Physics and Applied Mathematics, Columbia University, New York, NY 10027, USA}

\author{Y.J.~Uemura}
\affiliation{Department of Physics, Columbia University, New York, NY 10027, USA}

\author{R.~Khasanov}
\affiliation{Laboratory for Muon Spin Spectroscopy, Paul Scherrer Institute, CH-5232
Villigen PSI, Switzerland}

\begin{abstract}

High-pressure neutron powder diffraction, muon-spin rotation and magnetization studies of the structural, magnetic and the superconducting properties of the Ce-underdoped superconducting (SC) electron-doped cuprate system T$^{'}$-Pr$_{1.3-x}$La$_{0.7}$Ce$_{x}$CuO$_{4}$ with $x$ = 0.1 are reported. 
A strong reduction of the lattice constants $a$ and $c$ is observed under pressure. However, no indication of any pressure induced phase transition from T$^{'}$ to T structure is observed up to the maximum applied pressure of $p$ = 11 GPa. 
Large and non-linear increase of the short-range magnetic order temperature $T_{\rm so}$ in T$^{'}$-Pr$_{1.3-x}$La$_{0.7}$Ce$_{x}$CuO$_{4}$ ($x$ = 0.1) was observed under pressure. Simultaneously pressure causes a non-linear decrease of the SC transition temperature $T_{\rm c}$. All these experiments establish the short-range magnetic order as an intrinsic and a new competing phase in SC T$^{'}$-Pr$_{1.2}$La$_{0.7}$Ce$_{0.1}$CuO$_{4}$.  The observed pressure effects may be interpreted in terms of the improved nesting conditions through the reduction of the in-plane and out-of-plane lattice constants upon hydrostatic pressure.

\end{abstract}


\maketitle

\section{INTRODUCTION}

   One of the most important unresolved problems in high-transition temperature (high-$T_{\rm c}$) superconductors \cite{Bednorz} is the determination of the microscopic state when charge carriers, either holes \cite{Bednorz} or electrons \cite{Takagi,Tokura,Maple,Dalichaouch,Shengelaya,Tsukada}, are introduced to the CuO$_{2}$ planes of their insulating long-range antiferromagnetically ordered parent compounds \cite{Kastner}. One school of thought suggests that the doped charge carriers segregate into inhomogeneous patterns, such as stripes (spin and charge orders), to allow the antiferromagnetic (AFM) regions to survive \cite{Kivelson}. In this picture, the observed quasi two-dimensional (2D) incommensurate spin density wave (SDW) in hole-doped high-$T_{\rm c}$ superconductors, such as La$_{2-x}$Sr$_{x}$CuO$_{4}$ \cite{Tranquada1,Kimura,Vojta,Lake,Chang1,Guguchia1} and La$_{2}$CuO$_{4+\delta}$  \cite{Lee}, is due to remnants of the AFM insulating phase that compete with superconductivity (SC) \cite{Katano,Guguchia2016}. While charge ordering (CO) has emerged as a universal feature of hole-doped ($p$-type) cuprates \cite{Tranquada1,Kohsaka,Valla,Hucker,Chang,Julien,Ghiringelli}, observation of charge ordering in $n$-type cuprate system Nd$_{2-x}$Ce$_{x}$CuO$_{4}$ (NCCO) with $x$ = 0.14-0.15 was reported only very recently \cite{SilvaNeto,Sun}. 
Note that in $n$-type cuprates there are only limited number of comprehensive studies on SDW/CDW order and their interplay with superconductivity and structural properties\cite{SilvaNeto,Adachi2,Hasan,Kang,Greven,Armitage,Pascua}, underscoring a need for further experimental studies of n-type cuprates. 

Pr$_{1.3-x}$La$_{0.7}$Ce$_{x}$CuO$_{4}$ (PLCCO) is an n-type cuprate system that has recently been the focus of increased experimental investigation. The representations of two possible crystal structures, T$^{'}$ and T, for Pr$_{1.3-x}$La$_{0.7}$Ce$_{x}$CuO$_{4}$ are shown in Figs. 1a and b, respectively. The T$^{'}$-structure is  $I$4/$m$$m$$m$, 
Nd$_{2}$CuO$_{4}$-type and T structure is $I$4/$m$$m$$m$, K$_{2}$NiF$_{4}$ type. Cuprates with T$^{'}$ crystal structure have no apical oxygen O$_{ap}$ above or below the copper ions of the CuO$_{2}$-plane \cite{Armitage} as shown in Fig. 1a. The interesting characteristics of the electron-doped T$^{'}$-cuprates is that as-grown samples contain excess oxygen at the apical site. The apical oxygen is believed to induce disorder of the electrostatic potential in the CuO$_{2}$ plane, leading to the destruction of Cooper pairs \cite{Xu,Sekitani}. Therefore, completely removing the excess oxygen from as-grown samples is crucial to unveil intrinsic properties of the T$^{'}$-cuprates.
Interestingly, in T$^{'}$-Nd$_{2-x}$Ce$_{x}$CuO$_{4}$ thin films, superconductivity is achieved at any doping $x$ from 0.00 to 0.20 \cite{Matsumoto} by oxygen reduction annealing. This was claimed to be generic to T$^{'}$-cuprates with no apical oxygen. The role of the reduction process in the superconductivity of electron-doped cuprates has been a long-standing unsolved problem. Although the reduction annealing process is widely believed to involve removal of the apical oxygen, Raman, infrared transmission, and ultrasound studies on NCCO and PCCO suggest that the reduction process removes the oxygen in the CuO$_{2}$ plane \cite{Riou,Richard}  at high Cedoping ($x$ ${\textgreater}$ 0.1) and the apical oxygen at low Ce-doping. 
In this case, the in-plane oxygen defect created by the reduction is believed to be responsible for
superconductivity by destroying the long-range antiferromagnetism and increasing the
mobility of charge carriers. It was also shown  \cite{KangNature} that the microscopic process of oxygen reduction
repairs Cu deficiencies in the as-grown materials and creates oxygen vacancies in
the stoichiometric CuO$_{2}$ planes, effectively reducing disorder and providing
itinerant carriers for superconductivity. So far, there is no consensus as to which of the above mentioned scenarios is relevant to explain the reduction annealing process.

\begin{figure}[t!]
\centering
\includegraphics[width=1.55\linewidth]{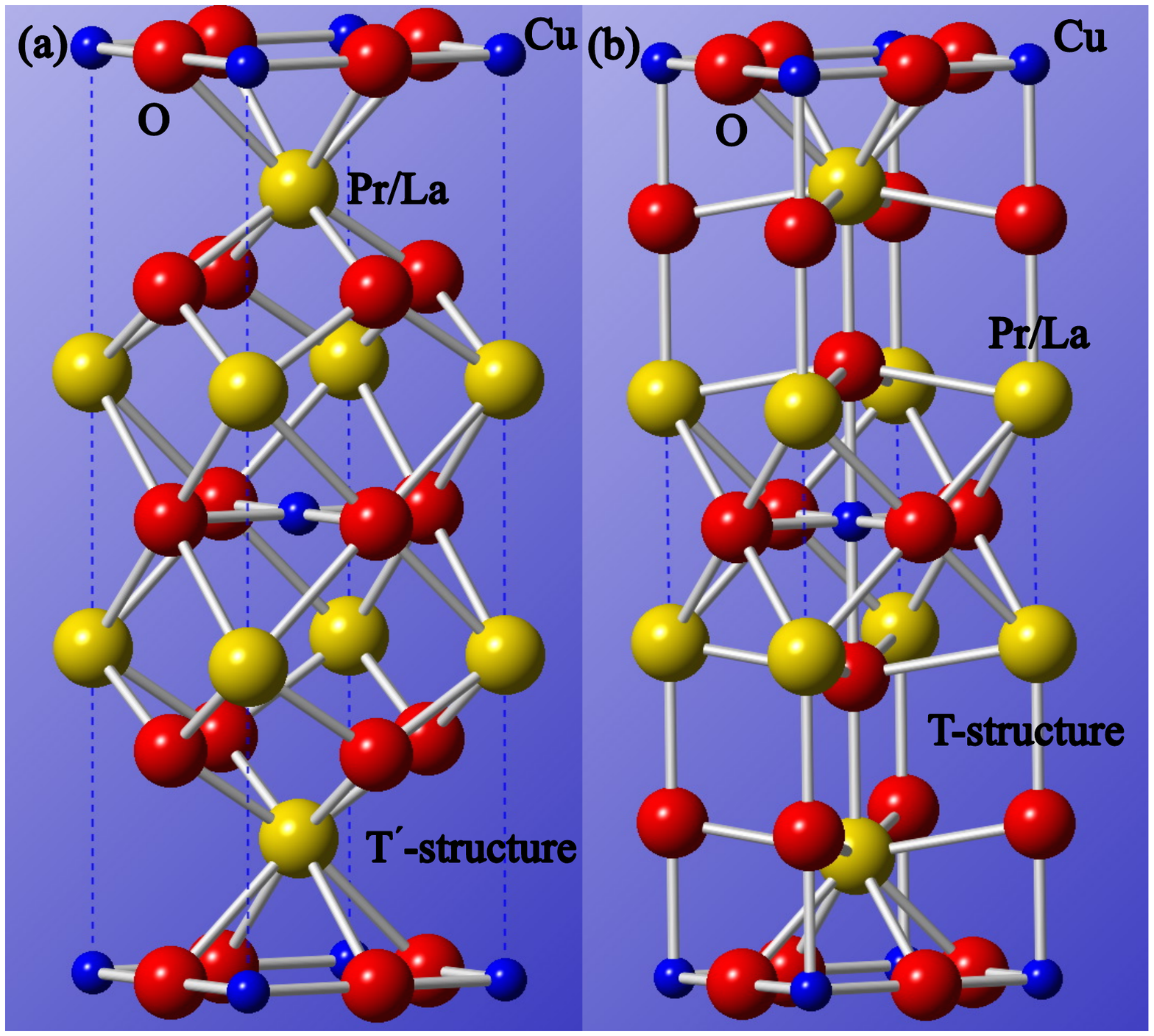}
\vspace{-1.3cm}
\caption{ (Color online) Representations of both T$^{'}$- (a) and T structures (b) of Pr$_{1.2}$La$_{0.7}$Ce$_{0.1}$CuO$_{4}$. T$^{'}$-structure is $I$4/$m$$m$$m$, Nd$_{2}$CuO$_{4}$-type and T-structure is $I$4/$m$$m$$m$, K$_{2}$NiF$_{4}$-type.}
\label{fig1}
\end{figure}

\begin{figure}[b!]
\centering
\includegraphics[width=1.0\linewidth]{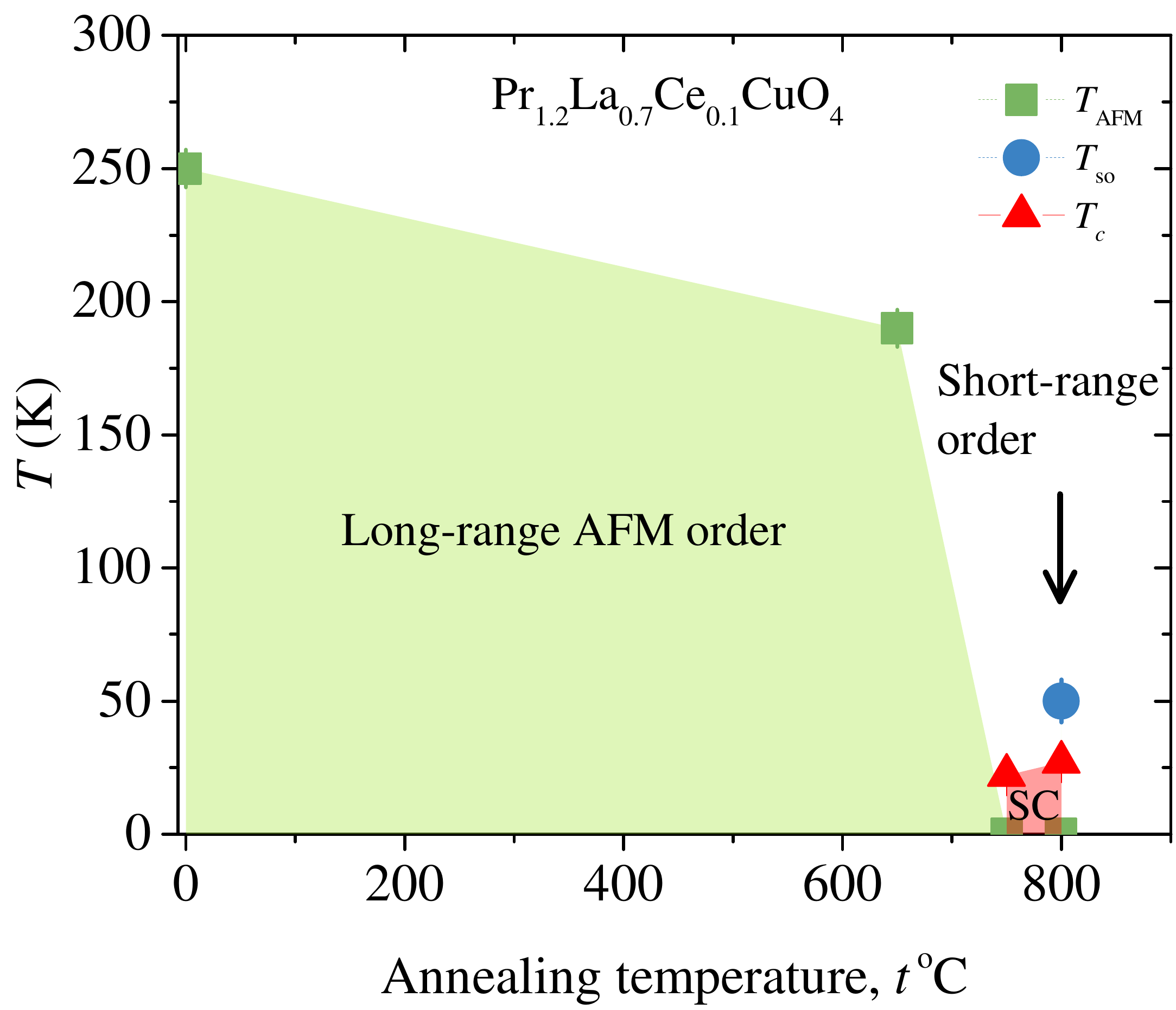}
\vspace{-0.7cm}
\caption{ (Color online) The long-range AFM, the short-range magnetic and the superconducting transition temperatures of Pr$_{1.2}$La$_{0.7}$Ce$_{0.1}$CuO$_{4}$ as a function of the sample annealing temperature in vacuum. The arrow indicates the sample, studied in the present work.}
\label{fig1}
\end{figure}

 Recent angle-resolved photoemission spectroscopy (ARPES) experiments \cite{Horio} in  SC single crystals of Ce-underdoped T$^{'}$-Pr$_{1.3-x}$La$_{0.7}$Ce$_{x}$CuO$_{4}$ ($x$ = 0.1) with $T_{\rm c}$ = 27 K, prepared through the improved reduction annealing \cite{Adachi1,Adachi2}, revealed the complete suppression of the AFM gap at the ''hot spots'' \cite{Horio}, i.e., the intersecting points of the paramagnetic Fermi surface and the antiferromagnetic Brillouin-zone boundary. This suggests that through improved annealing procedure, the long-range antiferromagnetic phase is removed from the underdoped regime (see Fig. 2). The question arises, whether the magnetic order is fully suppressed in improved annealed T$^{'}$-Pr$_{1.3-x}$La$_{0.7}$Ce$_{x}$CuO$_{4}$ or it exhibits short-range magnetic order, which can not be seen by ARPES, since it can not directly probe magnetism. In general, a fundamental question in electron-doped cuprates is whether SDW, CDW \cite{Sun} or stripe phases are emerging together with superconductivity as observed in the hole doped cuprates. Interestingly, the presence of short range magnetic order below ${\sim}$ 50 K in the SC T$^{'}$-Pr$_{1.2}$La$_{0.7}$Ce$_{0.1}$CuO$_{4}$ was demonstrated using muon-spin rotation (${\mu}$SR) experiments, implying that the reduction annealing gives rise to the change of the long-range AFM order to a short-range magnetic order \cite{Adachi2}. This is schematically demonstrated in Fig. 2, where we show the dependence of the magnetic and SC transition temperatures as a function of the sample annealing temperature.
Driving the long-range antiferromagnetic state to short-range magnetic state by annealing was also previously reported for Pr$_{0.88}$LaCe$_{0.12}$CuO$_{4}$ \cite{Wilson}.
Understanding the nature of this short-range magnetic order (i.e., whether it is intrinsic and related to the similar SDW, CDW, or stripe like order, as observed in hole-doped cuprates) and its relation to superconductivity is a main motivation of this work. 

   
   
  An important experimental parameter to tune the physical properties of the system is the hydrostatic pressure.
In $n$-type cuprates, only a limited number of pressure effects studies exist \cite{Markert,Ishiwata,Kamiyama,Wilhelm1998,Wilhelm2000,Rotundu}. 
To the best of our knowledge no pressure effect studies of the AFM order in $n$-type cuprates were reported so far. 

 In this paper, we report on the results of high pressure muon-spin rotation (${\mu}$SR) \cite{KhasanovPressure,Maisuradze}, neutron powder diffraction, and AC and DC susceptibility experiments for polycrystalline samples of the improved reduction annealed T$^{'}$-Pr$_{1.2}$La$_{0.7}$Ce$_{0.1}$CuO$_{4}$. Remarkably, substantial increase of the short-range magnetic order temperature $T_{\rm so}$ in Pr$_{1.2}$La$_{0.7}$Ce$_{0.1}$CuO$_{4}$ was observed, providing the first example of such a giant pressure effect on magnetism in electron-doped cuprates. We also found that the pressure effects have opposite signs for $T_{\rm so}$ and superconducting transition temperature $T_{\rm c}$, providing direct evidence for the competition between superconductivity and short-range magnetic order in this system. The pressure experiments strongly suggest that it is an intrinsic part of the phase diagram and is controlled by the Fermi surface properties of Pr$_{1.2}$La$_{0.7}$Ce$_{0.1}$CuO$_{4}$. Thus, a new competing phase to superconductivity in Pr$_{1.2}$La$_{0.7}$Ce$_{0.1}$CuO$_{4}$ has been demonstrated.  





\section{EXPERIMENTAL DETAILS}
\subsection{Sample preparation and characterization} 
The details of the synthesis of the polycrystalline samples of T$^{'}$-Pr$_{1.3-x}$La$_{0.7}$Ce$_{x}$CuO$_{4}$ were reported previously \cite{Adachi1,Adachi2}. 
${\mu}$SR experiments under pressure were performed at the
GPD instrument (${\mu}$SRE1 beamline) of the Paul Scherrer
Institute (Villigen, Switzerland). The low background
GPS (${\pi}$M3 beamline) instrument was used to study the system T$^{'}$-Pr$_{1.3-x}$La$_{0.7}$Ce$_{x}$CuO$_{4}$ at ambient pressure. The pressure effects on the structural properties of T$^{'}$-Pr$_{1.3-x}$La$_{0.7}$Ce$_{x}$CuO$_{4}$ were studied using time-of-flight neutron powder diffraction at SNAP the high pressure diffractometer (BL-3) at  the  Spallation  Neutron  Source (SNS) of the Oak Ridge National Laboratory. All measurements reported here were performed on samples from the same batch.

\subsection{Pressure cells for ${\mu}$SR, neutron powder diffraction and magnetization experiments}

Pressures up to 2.3 GPa were generated in a double wall piston-cylinder type of cell made of MP35N material,
especially designed to perform ${\mu}$SR experiments under pressure
\cite{KhasanovPressure,Maisuradze}. As a pressure transmitting medium Daphne
oil was used. The pressure was measured by tracking the SC transition
of a very small indium plate by AC susceptibility. The amount of the sample in the pressure cell was optimized and the fraction of the muons in the sample was approximately 40 ${\%}$. 

In neutron powder diffraction experiments, pressures up to 11 GPa were generated in a Diamond anvil cell (DAC), using large synthetic diamonds and a new anvil design based on conical anvil support \cite{Boehler1}, a new generation of large cell developed at the SNAP beamline \cite{Boehler2}. 
The SNAP DAC was pressurized using an integrated gas driven membrane set-up.

 AC susceptibility measurements were performed by using a home made AC magnetometer with a measuring field 
${\mu}_{0}$$H_{\rm AC}$ ${\sim}$ 0.1 mT and frequency ${\nu}$ = 96 Hz.  
In order to keep the position of the sample unchanged during the series of AC susceptibility under pressure measurements, 
the excitation and the two pick-up coils were wound directly on the cell.
Note that single-phase lock-in amplifier is used, which allows to measure either in-phase component (the real part of susceptibility) or out-of-phase component (the imaginary part of susceptibility). For this particular sample, the imaginary part of AC susceptibility was measured.

 The DC susceptibility was measured under pressures up to 2.5 GPa
by a SQUID magnetometer ($Quantum$ $Design$ MPMS-XL). 
Pressures were generated using a diamond anvil cell (DAC) \cite{Giriat} 
filled with Daphne oil which served as a pressure-transmitting medium. 
The pressure at low temperatures was determined 
by detecting the pressure dependence of the SC transition temperature of Pb.

\begin{figure}[t!]
\centering
\includegraphics[width=1.0\linewidth]{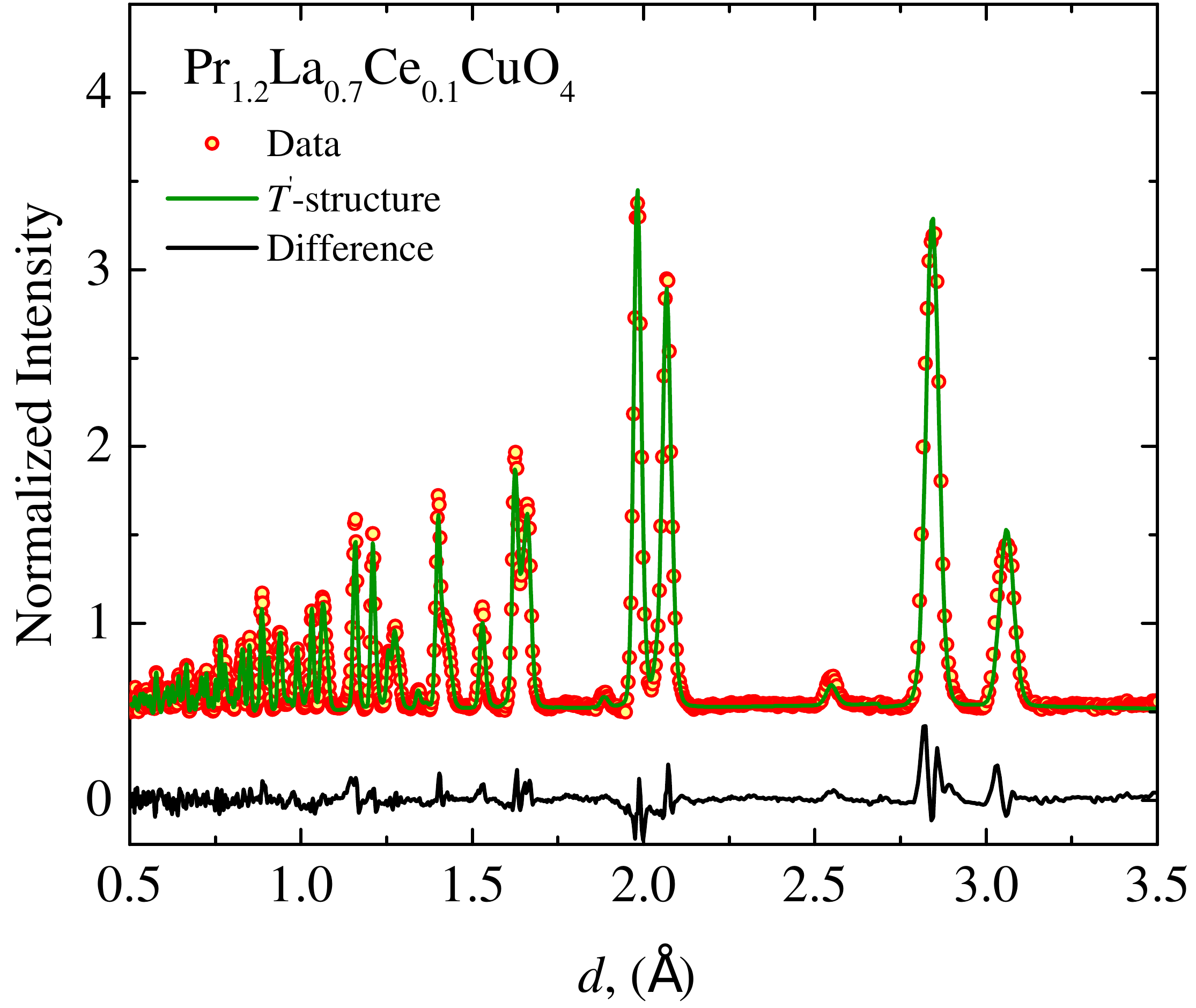}
\vspace{-0.7cm}
\caption{ (Color online) Rietveld refinement of the neutron diffraction pattern for the T$^{'}$-Pr$_{1.2}$La$_{0.7}$Ce$_{0.1}$CuO$_{4}$ at 300 K at ambient conditions. The observed, calculated, and difference
plots are shown by solid circles, dark green solid line, and light green
solid line, respectively. Bragg reflections are compatible
with space group $I$4/$m$$m$$m$, Nd$_{2}$CuO$_{4}$-type.}
\label{fig1}
\end{figure}

\begin{figure*}[t!]
\centering
\includegraphics[width=1.0\linewidth]{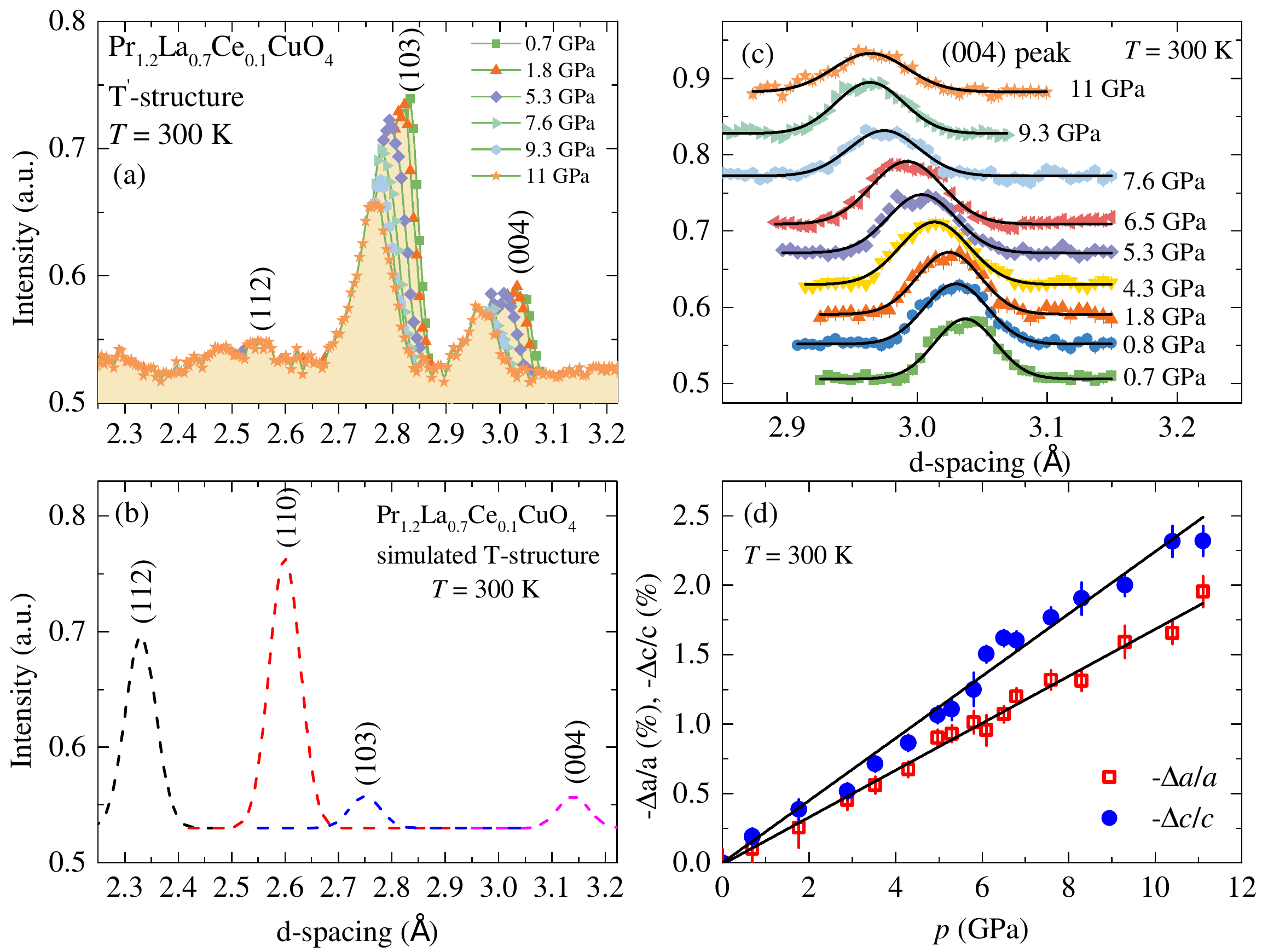}
\vspace{-0.7cm}
\caption{ (Color online) (a) The (112), (103), and (004) Bragg peaks of Pr$_{1.2}$La$_{0.7}$Ce$_{0.1}$CuO$_{4}$, recorded at various pressures up to $p$ = 11 GPa. The data are corresponding to $T^{'}$ structure. (b) Simulated Bragg peaks for the T-structure of Pr$_{1.2}$La$_{0.7}$Ce$_{0.1}$CuO$_{4}$, shown in the same d-spacing range as in panel (a). (c) The (004) Bragg reflection, recorded at 300 K at various hydrostatic pressures, illustrating the continuous shift of the $c$-axis peak towards the lower $d$-spacing with increasing the pressure. The solid lines represent the Gaussian fits to the data. (d) Pressure dependence of the relative pressure shifts ${\Delta}$$a$/$a$ and ${\Delta}$$c$/$c$ for the lattice parameters $a$ and $c$ of Pr$_{1.2}$La$_{0.7}$Ce$_{0.1}$CuO$_{4}$ at $T$ = 300 K. The solid lines are linear fits of the data.}
\label{fig1}
\end{figure*}

\begin{figure}[t!]
\centering
\includegraphics[width=1.0\linewidth]{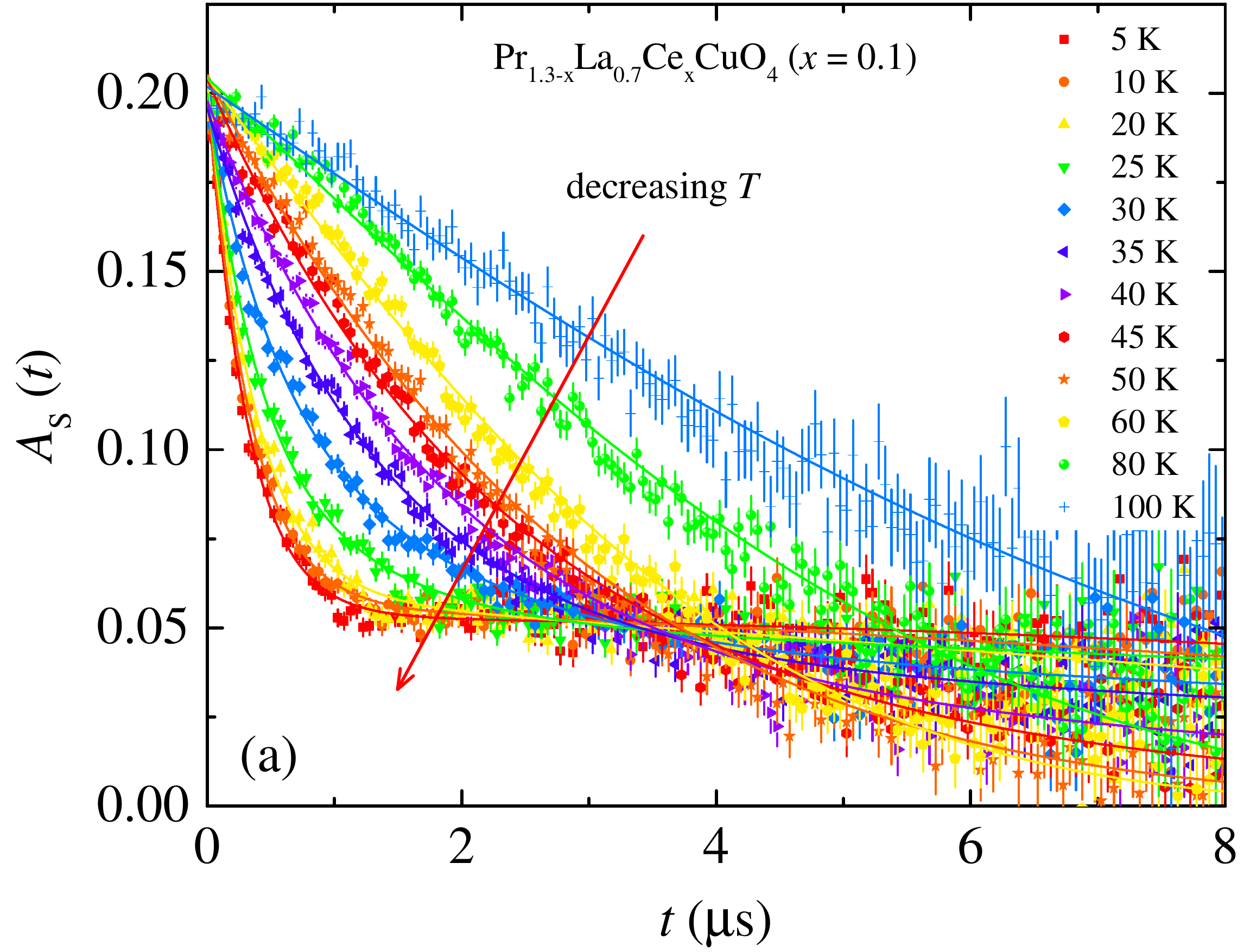}
\includegraphics[width=1.0\linewidth]{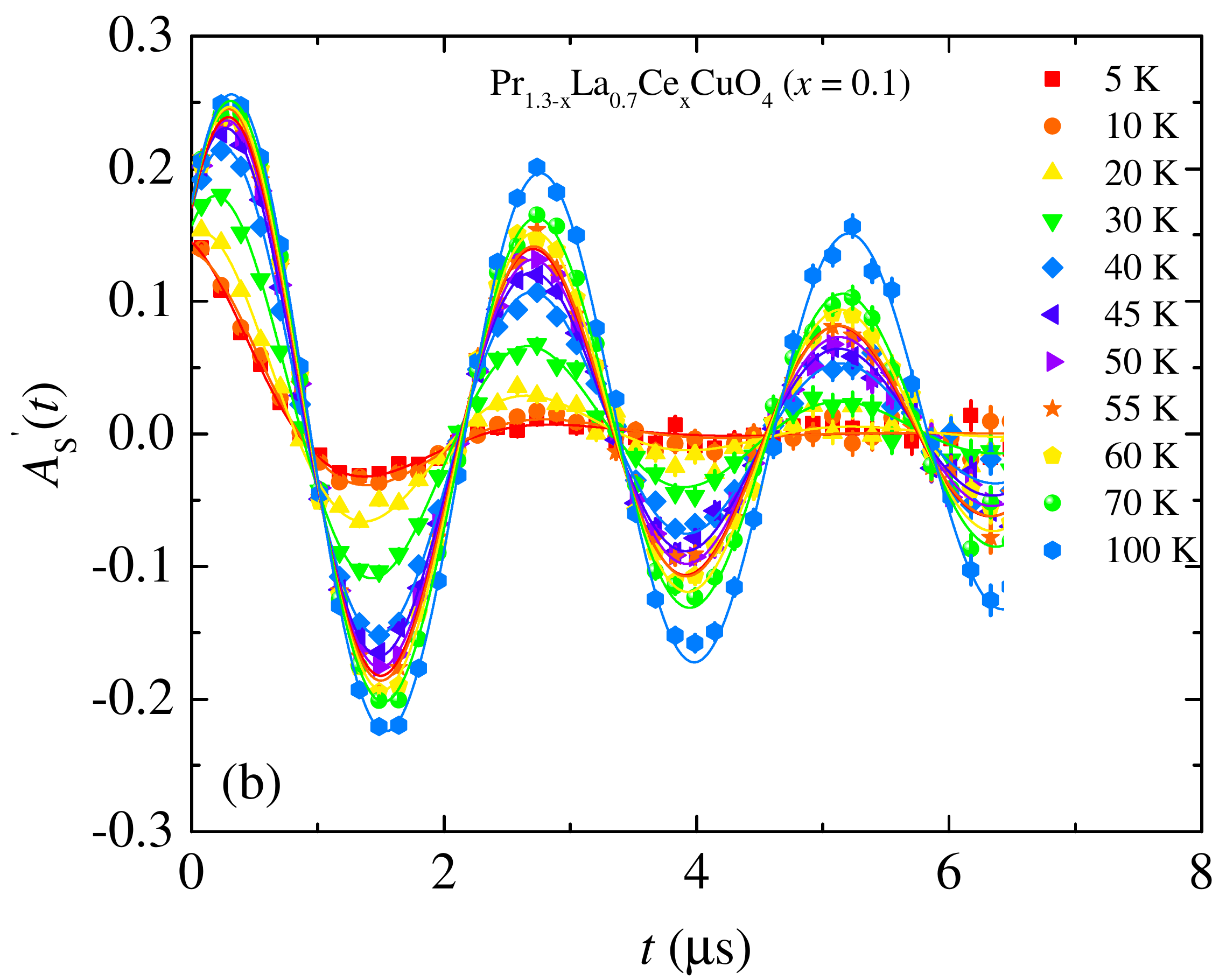}
\vspace{-0.7cm}
\caption{ (Color online) ZF ${\mu}$SR (a) and WTF ${\mu}$SR (b)  time spectra for Pr$_{1.2}$La$_{0.7}$Ce$_{0.1}$CuO$_{4}$ recorded at various temperatures without the pressure cell. The solid lines represent fits to the data by means of Eq.~(2) and Eq.~(3).}
\label{fig1}
\end{figure}

\begin{figure*}[t!]
\centering
\includegraphics[width=1.0\linewidth]{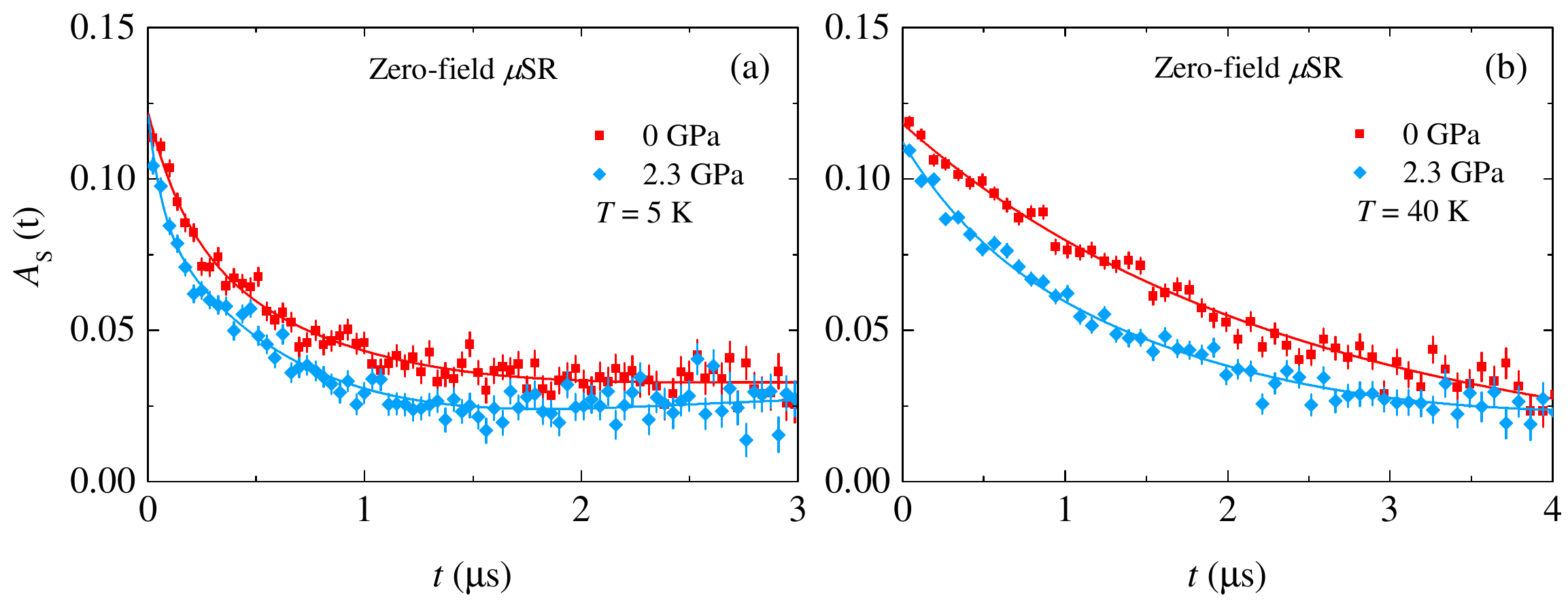}
\vspace{-0.7cm}
\caption{ (Color online) ZF ${\mu}$SR time spectra $A_{S}$ for Pr$_{1.2}$La$_{0.7}$Ce$_{0.1}$CuO$_{4}$ recorded at $T$ = 5 K (a) and 40 K (b) for ambient and maximum applied pressure of $p$ = 2.3 GPa. The spectra are shown after substraction  of the background signal from the pressure cell. The solid lines represent fits to the data by means of Eq.~(2).}
\label{fig1}
\end{figure*}

\begin{figure}[b!]
\includegraphics[width=1.0\linewidth]{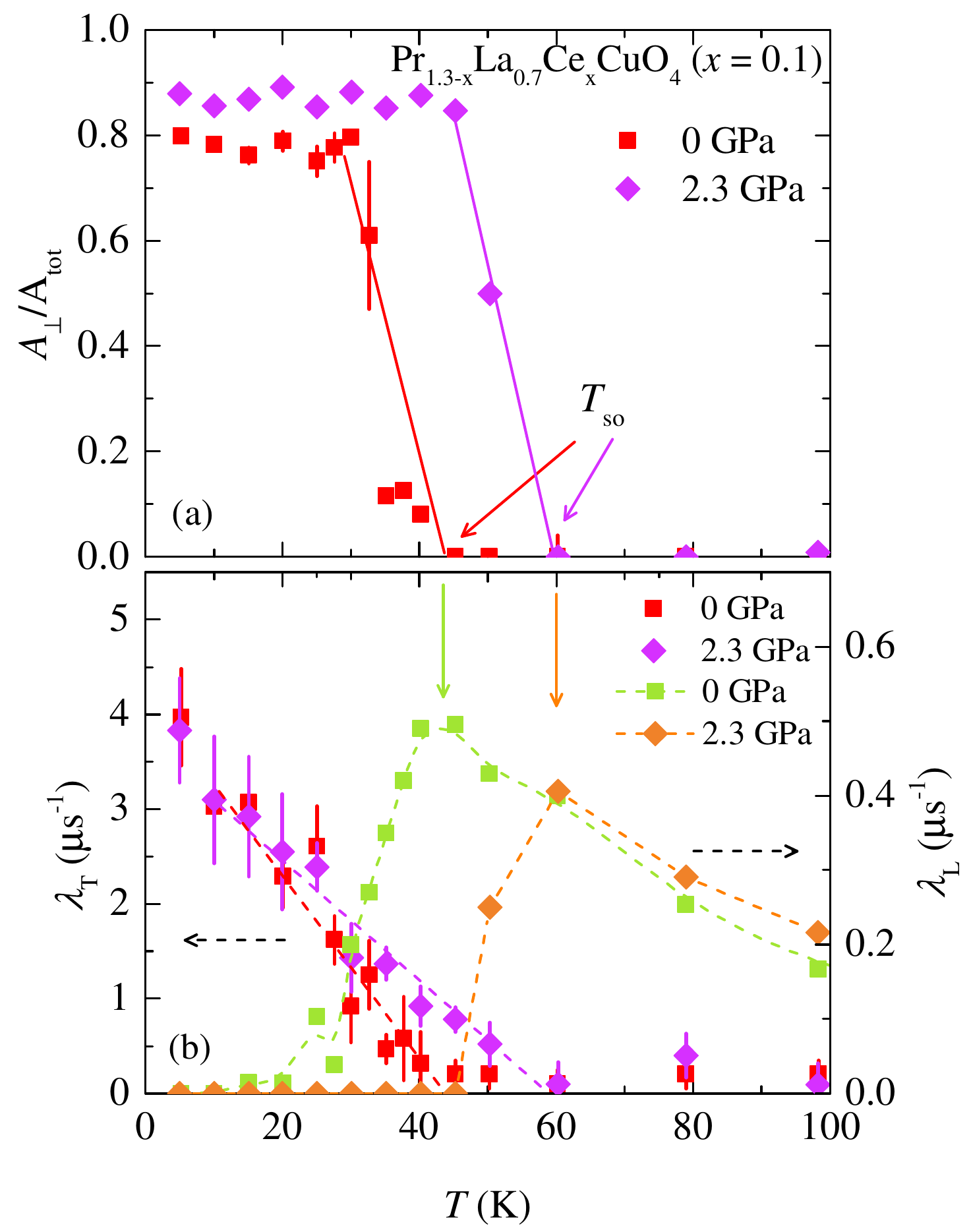}
\vspace{-0.7cm}
\caption{(Color online) The temperature dependence of the ratio $A_{\perp}$/$A_{tot}$ (a) and the relaxation rates ${\lambda}_{T}$, ${\lambda}_{L}$ (b) of Pr$_{1.2}$La$_{0.7}$Ce$_{0.1}$CuO$_{4}$, recorded at $p$ = 0 GPa and $p$ = 2.3 GPa. The arrows denote the static magnetic order temperature $T_{\rm so}$. The solid lines are guides to the eye.}  
\label{fig1}
\end{figure}

\section{RESULTS}

\subsection{High pressure Neutron Powder Diffraction Experiments}

 Previous high-pressure X-ray study up to 0.6 GPa in undoped Nd$_{2}$CuO$_{4}$ \cite{Kamiyama}  and optimally doped Nd$_{1.835}$Ce$_{0.165}$CuO$_{4}$ showed a decrease of the lattice parameters under pressure. Higher pressure experiments showed that in the parent compounds Nd$_{2}$CuO$_{4}$ \cite{Wilhelm1998} and Pr$_{2}$CuO$_{4}$ \cite{Wilhelm2000} a T$^{'}$ to $T$ structural transition takes place at 21.5 GPa and 15.1 GPa, respectively. Recently, it was shown for Nd$_{1.835}$Ce$_{0.165}$CuO$_{4}$ \cite{Rotundu} that the transformation of T$^{'}$ to $T$-phase  takes place at 2.7 GPa, i.e., at much lower pressure than for the parent compounds. This implies that the knowledge of the structural details of electron-doped cuprates under pressure are important in better understanding the pressure evolution of  magnetic and superconducting properties. Below we show Neutron diffraction data for the present Ce-underdoped electron-doped cuprate T$^{'}$-Pr$_{1.2}$La$_{0.7}$Ce$_{0.1}$CuO$_{4}$ under pressure as high as $p$ = 11 GPa. 

The crystal structure of Pr$_{1.2}$La$_{0.7}$Ce$_{0.1}$CuO$_{4}$ 
at room temperature (RT) (see Fig. 3) was well refined through Rietveld refinements to the raw
neutron diffraction data using the program GSAS-II \cite{TobyBH}, employing the tetragonal $I$4/$m$$m$$m$ space group (Nd$_{2}$CuO$_{4}$-type).  An example of the refinement profile is shown in Fig. 3. The weighted fit residual is Rwp ${\sim}$ 2.8 ${\%}$, indicating a fairly good refinement.
No obvious secondary phase, including the $T$-structural phase, can be detected, pointing 
to a pure T$^{'}$ phase in the studied Pr$_{1.2}$La$_{0.7}$Ce$_{0.1}$CuO$_{4}$ sample.
Figure 4a shows the evolution of the (112), (103) and (004) Bragg peaks from the T$^{'}$ structure of Pr$_{1.2}$La$_{0.7}$Ce$_{0.1}$CuO$_{4}$ measured at various pressures up to p=11 GPa. A systematic shift of the peaks towards lower $d$-spacing is observed under pressure, consistent with a monotonic reduction in the lattice constants. However, up to the maximum applied pressure of $p$ = 11 GPa, there is no indication of the pressure induced phase transition from T$^{'}$ to T structure, in which case the high intensity Bragg peaks (112) and (110) as well as the  much weaker (103) and (004) peaks are expected, as demonstrated by the simulated Bragg peaks from the $T$ structure in Fig. 4b. This points to the robustness of the T$^{'}$ structure in Pr$_{1.2}$La$_{0.7}$Ce$_{0.1}$CuO$_{4}$ in the investigated pressure range. The (004) Bragg peaks are shown separately in Fig. 4c. By fitting with a Gaussian function, drawn with the black solid lines through the data, the lattice constant $c$ was estimated at various pressures. The pressure dependence of its relative shift -${\Delta}$$c$/$c$ = ($c(0)$-$c(p)$)/$c(0)$ is shown in Fig. 4d, showing the continuous decrease of the $c$-axis under pressure. Using the value of $c$ and the d-spacing of the (103) Bragg peak, the lattice constant $a$ was obtained. The pressure evolution of the relative change in the c-axis -${\Delta}$$a$/$a$ is shown in Fig. 4d, revealing a similar linear pressure induced reduction of the $a$-axis. These results point to a nearly isotropic compression of the lattice in T$^{'}$-Pr$_{1.2}$La$_{0.7}$Ce$_{0.1}$CuO$_{4}$.

\subsection{Zero-field (ZF) and Weak-transverse field (WTF) ${\mu}$SR experiments}

   Figure~5a shows representative ZF ${\mu}$SR time spectra for polycrystalline Pr$_{1.2}$La$_{0.7}$Ce$_{0.1}$CuO$_{4}$, recorded at various temperatures. 
At high temperatures, only a very weak depolarization of the ${\mu}$SR
signal is observed. This weak depolarization reflect the occurrence of a small Gaussian Kubo-Toyabe
depolarization, originating from the interaction of the muon spin with randomly oriented nuclear
magnetic moments. In addition, small Pr$^{3+}$ moments, which are dense and randomly distributed also contribute to this relaxation \cite{Kadono,Risdiana}.
Upon lowering the temperature the relaxation rate of the ${\mu}$SR signal increases due to the development of the Cu-spin correlation.
No muon-spin precession is observed even at the lowest temperature $T$ = 5 K and only a rapidly depolarizing ${\mu}$SR signal is observed.
The fast depolarization of the ${\mu}$SR signal (with no trace of an oscillation) is due to a broad distribution of static fields.
This is supported by the fact that ZF ${\mu}$SR signal observed at temperatures below 50 K shows the recovery of asymmetry at longer times (see Fig. 5a), which is typical for static magnetically ordered systems. Static short-range magnetic order in the single crystal of Pr$_{1.2}$La$_{0.7}$Ce$_{0.1}$CuO$_{4}$ was previously established by ZF- ${\mu}$SR and longitudinal field (LF) ${\mu}$SR experiments  \cite{Adachi2}. In the following, we present   
how the magnetic and the SC properties of Pr$_{1.2}$La$_{0.7}$Ce$_{0.1}$CuO$_{4}$ evolve with hydrostatic pressure.

 Figs. 6a and b show the ambient and highest pressure ($p$ = 2.3 GPa) ZF ${\mu}$SR spectra of Pr$_{1.2}$La$_{0.7}$Ce$_{0.1}$CuO$_{4}$, recorded at 5 K and 40 K, respectively. It is clear that upon application of pressure magnetic response is enhanced both at 5 K and 40 K. 
To get the quantitative information about the pressure effects, the ${\mu}$SR data in the whole temperature range were analyzed by
decomposing the signal into a contribution of the sample and a contribution of the pressure cell, since in the high pressure ${\mu}$SR experiments, a substantial fraction of the ${\mu}$SR asymmetry originates from muons stopping in the MP35N pressure cell surrounding the sample:
\begin{equation}
A(t)=A_S(t)+A_{PC}(t),
\end{equation}
where $A_{S}(t)$ and $A_{PC}(t)$ are the asymmetries, belonging to the sample and the pressure cell, respectively.
The pressure cell signal was analyzed by a damped Kubo-Toyabe function \cite{KhasanovPressure,Maisuradze}.
The response of the sample (see Fig. 5a) is analysed by the following function \cite{GuguchiaJSNM,Maeter}: 
\begin{equation}
A_S(t)=\Bigg[A_{\perp}e^{-\lambda_{T}t}+A_{\parallel}e^{-\lambda_{L}t}\Bigg]e^{-\sigma_{N}^2t^2/2}.
\label{eq1}
\end{equation}
The ratio $A_{\perp}/A_{\parallel}$ (the ratio of the transversal and the longitudinal relaxing components of the asymmetry signal) reveals the average degree to
which $P_{\mu}$ aligns with the local field. In a fully magnetic polycrystalline sample, with a static component to the local field, because the crystallites orient randomly with respect to $P_{\mu}$, $A_{\perp}$/($A_{\perp}$ + $A_{\parallel}$) = $\frac{2}{3}$, the isotropic average perpendicular component of the muon spin ${\lambda_T}$ and ${\lambda_L}$ are the relaxation rates characterising the damping of the transversal and the longitudinal components of the ${\mu}$SR signal, respectively. 
The transversal relaxation rate ${\lambda_T}$ is a measure of the width of the static magnetic field distribution at the muon site, and also reflects dynamical effects
(spin fluctuations). The longitudinal relaxation rate ${\lambda_L}$ is determined by dynamic magnetic fluctuations only \cite{GuguchiaJSNM,Maeter}.
${\sigma_{N}}$ is the Gaussian relaxation rate, caused by the nuclear spins and the small Pr$^{3+}$ moments. 
The total initial assymetry $A_{\rm tot}$ = $A_{\rm S}$(0) + $A_{\rm PC}$(0) ${\simeq}$ 0.28 is a temperature independent constant. 
A typical fraction of muons stopped in the sample was $A_{\rm S}$(0)/$A_{\rm tot}$ ${\simeq}$ 0.40(3) which was assumed to be temperature independent in the analysis. The ${\mu}$SR time spectra were analyzed using the free software package MUSRFIT \cite{AndreasSuter}.

 Figure 7 summarises the results of the above analysis. Namely, in Figs. 7a and b,  we show the temperature dependences of the ratio $A_{\perp}$/$A_{tot}$ ($A_{tot}$ = $A_{\perp}$ + $A_{\parallel}$) and the transverse relaxation rate ${\lambda}_{T}$ as well as the longitudinal relaxation rate ${\lambda}_{L}$, respectively, for Pr$_{1.2}$La$_{0.7}$Ce$_{0.1}$CuO$_{4}$, measured at $p$ = 0 GPa and 2.3 GPa. $A_{\perp}$ and $A_{\parallel}$ are the fractions of the transversal and the longitudinal relaxing components of the asymmetry signal, respectively. In the present case, $A_{\perp}$/$A_{tot}$ starts to increase below $T_{\rm so}$ ${\simeq}$ 45 K and reaches the value of $V_{\rm m}$  ${\simeq}$ 0.8 at the base temperature. 
Note that the obtained value $A_{\perp}$/$A_{tot}$ ${\simeq}$ 0.8 is higher than the one $\frac{2}{3}$, expected for the polycrystalline sample. This is may be related to some preferred orientation in our sample. Since Pr$_{1.2}$La$_{0.7}$Ce$_{0.1}$CuO$_{4}$ has a two-dimensional structure (actually, the single crystal always grows in the direction parallel to the CuO$_{2}$ plane), even in powderized sintered pellet, there might be some preferred orientation. On the other hand, this could also be related to the existence of some fraction of fluctuating Cu spins, which was previously discussed \cite{Adachi2}. The high value of $A_{\perp}$/$A_{tot}$ indicates that the short-range magnetic order occupies the nearly whole volume of this superconducting Pr$_{1.2}$La$_{0.7}$Ce$_{0.1}$CuO$_{4}$. ${\lambda}_{T}$, characterizing the distribution of local fields shows monotonous increase with decreasing the temperature below $T_{\rm so}$ ${\simeq}$ 45 K.  ${\lambda}_{L}$, characterizing the muon spin relaxation due to fluctuating magnetic fields also shows a clear peak at $T_{\rm so}$, which is typical for the magnetic phase transition. Note that in the ordered state the values of ${\lambda}_{L}$ is almost zero, consistent with the presence of dominant static magnetic order in Pr$_{1.2}$La$_{0.7}$Ce$_{0.1}$CuO$_{4}$. 
It is evident from Fig. 7a that the magnetic fraction reaches 100 ${\%}$ at low $T$ for both pressures. 
In addition, the magnetic order temperature $T_{\rm so}$ in Pr$_{1.2}$La$_{0.7}$Ce$_{0.1}$CuO$_{4}$ increases substantially, by ${\sim}$ 15 K, at the highest applied pressure of $p$ ${\simeq}$ 2.3 GPa. The pressure induced enhancement of  $T_{\rm so}$ can also be seen from the temperature dependences of ${\lambda}_{T}$ and ${\lambda}_{L}$, shown in Fig. 7b.
 
\begin{figure}[t!]
\includegraphics[width=1.0\linewidth]{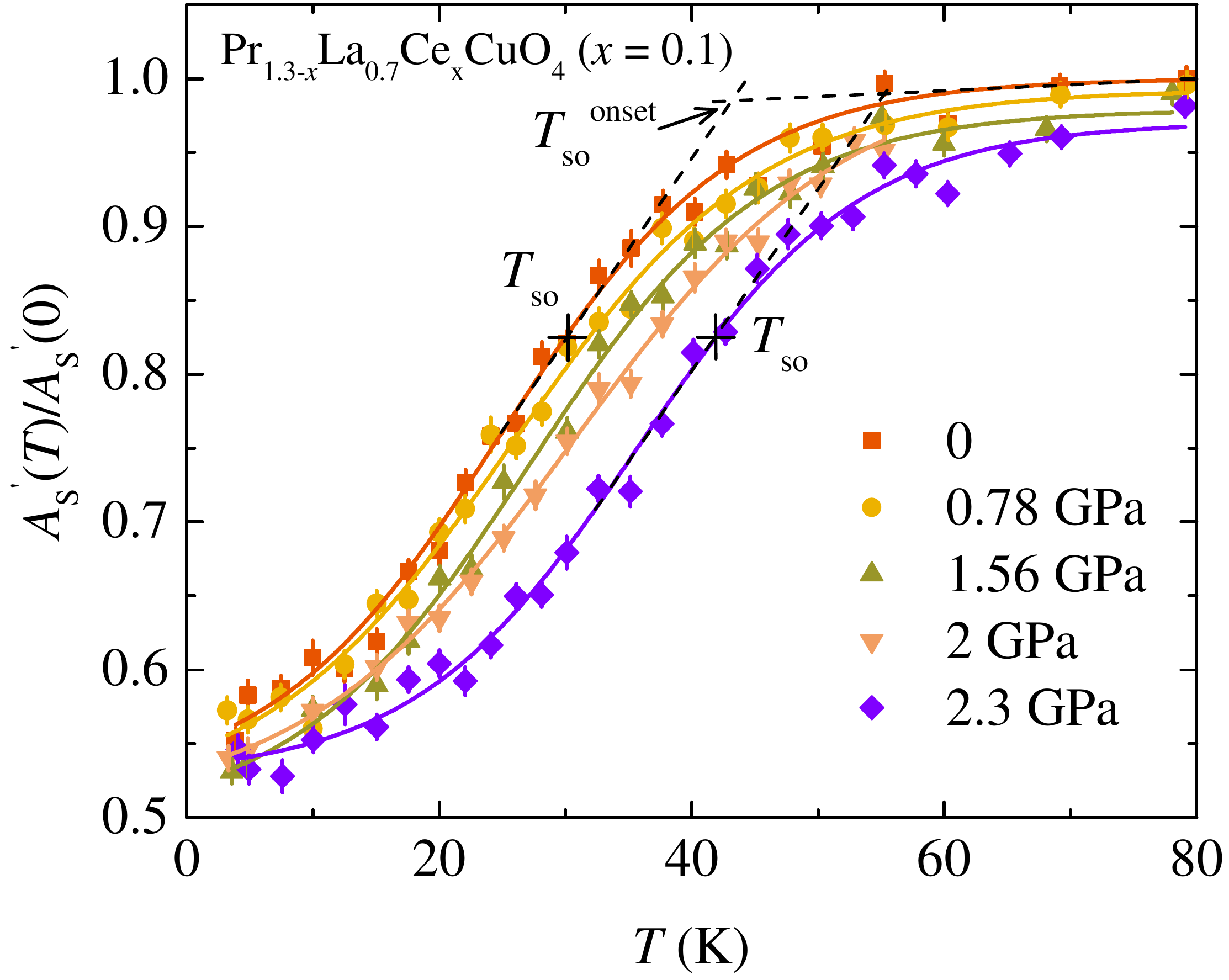}
\vspace{-0.7cm}
\caption{ (Color online) The temperature dependence of the normalised TF asymmetry $A_{S}^{'}($T$)/A_{S}^{'}(0)$ for Pr$_{1.2}$La$_{0.7}$Ce$_{0.1}$CuO$_{4}$, recorded at various pressures up to $p$ = 2.3 GPa. The crosses mark the magnetic order temperatures $T_{\rm so}$ for $p$ = 0 and 2.3 GPa. The solid
lines represent fits to the data by means of Eq. (1).}  
\label{fig1}
\end{figure}

   In order to gain more insight into the above short-range magnetic transition, TF-${\mu}$SR experiments in weak transverse field were carried out. Figure~5b shows representative Weak-transverse field (WTF) ${\mu}$SR time spectra for polycrystalline Pr$_{1.2}$La$_{0.7}$Ce$_{0.1}$CuO$_{4}$, recorded at various temperatures. This clearly shows the reduction of the amplitude of the ${\mu}$SR signal upon lowering the temperature below ${\sim}$ 50 K, indicating the appearance of the magnetic order. The WTF-${\mu}$SR spectra, shown in Fig. 5b were fitted in the time domain with a slowly relaxing signal with the precession frequency corresponding to the applied field of $\mu_{0}H$ = 3 mT:

\begin{equation}
\begin{split}
A^{'}(t)= A_S^{'}(t)+A_{PC}^{'}(t) =  \\   (A_{S}^{'}e^{-\lambda^{'} t}  +
A_{PC}^{'}e^{-\lambda_{PC}^{'}t}){\cos}(\gamma_{\mu}Bt + {\varphi})
\label{eq1}
\end{split}
\end{equation}

 where, $t$ is time after muon implantation, ${\varphi}$ is the phase, $A^{'}$(t) is the time-dependent asymmetry, and $\gamma_{\mu}/(2{\pi}) \simeq 135.5$~MHz/T is the muon gyromagnetic ratio. $A_{S}^{'}$ is the amplitude of the oscillating component (related to the paramagnetic volume fraction of the sample). ${\lambda}^{'}$  is an exponential damping rate due to paramagnetic spin fluctuations and nuclear dipolar moments. $B$ is the applied magnetic field, experienced by the muons stopped in the paramagnetic part of the sample. From these refinements, the paramagnetic volume fraction at each temperature $T$ was estimated as $A_{S}^{'}(T)$/$A_{S}^{'}(0)$, where $A_{S}^{'}(0)$ is the amplitude in the paramagnetic phase at high temperature. $A_{PC}^{'}$ and ${\lambda}_{PC}^{'}$ is the amplitude and the relaxation rate of the pressure cell signal.
Pressure cell amplitude was estimated to be $A_{PC}^{'}$ = 0.17 and it was kept constant as a function of temperature in the analysis. Note that the data, shown in Figs. 5a and b are recorded without the pressure cell ($A_{PC}^{'}$ = 0).

 Figure 8 shows the normalised TF-${\mu}$SR asymmetry $A_s^{'}(T)/A_s^{'}(0)$ for Pr$_{1.2}$La$_{0.7}$Ce$_{0.1}$CuO$_{4}$ extracted from the ${\mu}$SR spectra, shown in Fig. 5b (following the Eq. 3), as a function of temperature for ambient and selected applied pressures in an applied field of ${\mu}_{0}$$H$ = 3 mT. For $p$ = 0 GPa and $T$ ${\textgreater}$  50 K, $A_s^{'}(T)/A_s^{'}(0)$ saturates nearly at a maximum value, indicating that
nearly the whole sample is in the paramagnetic state, and all the muon spins precess in the applied magnetic field.
Below 50 K, $A_s^{'}(T)/A_s^{'}(0)$ continuously decreases with decreasing temperature. 
The reduction of $A_s^{'}(T)/A_s^{'}(0)$ signals the appearance of magnetic order,
where the muon spins experience a local magnetic field larger than the applied magnetic field.
As a result, the fraction of muons in the paramagnetic state decreases.
The onset temperature $T_{\rm so}^{onset}$ is defined as the temperatures where the linearly extrapolated low and high temperature data points intersect.
The midpoint of the transition ($T_{\rm so}$) was determined by using the phenomenological function \cite{GuguchiaPRL,AndreasSuter}:  
\begin{equation}
A_{S}^{'}(T)/A_{S}^{'}(0) = a\Bigg[1-\frac{1}{{\exp}[(T-T_{so})/{\Delta}T_{so}]+1}\Bigg]+b,
\label{eq1}
\end{equation}
${\Delta}T_{\rm so}$ is the width of the transition, whereas $a$ and $b$ are empirical parameters. 
Analyzing the $p$ = 0 GPa data in Fig.~8 with Eq.~(4) yields:
$T_{\rm so}$ = 30(1) K, which indicated by cross in Fig. 8. 
Remarkably, a strong and non-linear increase of $T_{\rm so}$ in Pr$_{1.2}$La$_{0.7}$Ce$_{0.1}$CuO$_{4}$ is observed under pressure.

\subsection{High pressure magnetic susceptibility data}

\begin{figure}[t!]
\centering
\includegraphics[width=1.0\linewidth]{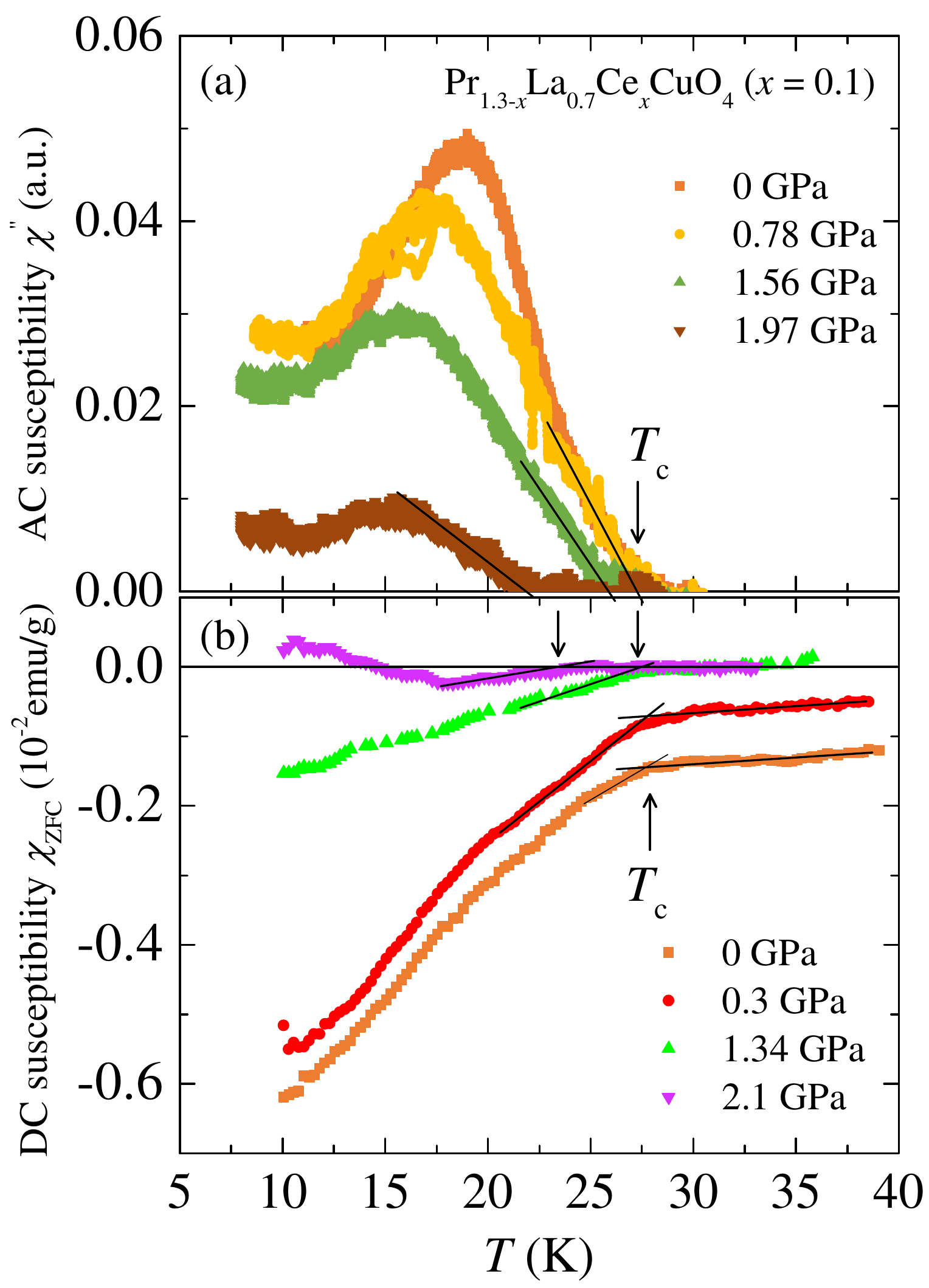}
\vspace{-0.7cm}
\caption{ (Color online) Temperature dependence of the AC (a) and DC (b) diamagnetic susceptibilities, $\chi^{''}$ and $\chi_{\rm ZFC}$, respectively, of Pr$_{1.2}$La$_{0.7}$Ce$_{0.1}$CuO$_{4}$, measured at ambient and at various applied hydrostatic pressures. The arrows denote the superconducting transition temperature $T_{\rm c}$. For clarity, the data for $p$ = 0 and 0.3 GPa are shifted vertically down from the zero line.} 
\label{fig1}
\end{figure}

 Figure 9a and b show the temperature dependences of AC and DC magnetic susceptibilities $\chi^{''}$ and $\chi_{\rm ZFC}$, respectively, for Pr$_{1.2}$La$_{0.7}$Ce$_{0.1}$CuO$_{4}$ at ambient and at various applied pressures up to $p$ ${\simeq}$ 2.3 GPa, after substraction of the background signal from the empty pressure cell. Note that zero-field-cooled (ZFC)  DC susceptibility $\chi_{\rm ZFC}$ was measured in a magnetic field of ${\mu}_{0}H$ = 0.5 mT using the DAC, where only very tiny amount of sample is used.  On the other hand, $\chi^{''}$ was measured using exactly the same sample and exactly the same cell as the one used for the ${\mu}$SR experiments, which requires a large amount of sample. 
At $p$ = 0 GPa, the sample Pr$_{1.2}$La$_{0.7}$Ce$_{0.1}$CuO$_{4}$ shows superconductivity with $T_{\rm c}$ ${\simeq}$ 27 K. The magnitude of  $\chi_{\rm ZFC}$ = -0.49(3) ${\times}$ 10$^{-2}$ (emu/g) at $T$ = 10 K, providing a lower limit of 49(3) ${\%}$ for the SC volume fraction. This implies the bulk character of superconductivity in the sample. $T_{\rm c}$ shows the modest decrease with increasing pressure, i.e., it decreases from $T_{\rm c}$ ${\simeq}$ 27 K to $T_{\rm c}$ ${\simeq}$ 23 K at $p$ ${\simeq}$ 2.1 GPa. On the other hand, application of pressure leads to the substantial decrease of the diamagnetic response, resulting in difficulties of observing the SC response above $p$ ${\simeq}$ 2.1 GPa. It is difficult to conclude whether superconductivity in Pr$_{1.2}$La$_{0.7}$Ce$_{0.1}$CuO$_{4}$ is either fully suppressed above $p$ ${\simeq}$ 2.1 GPa or becomes filamentary.   

\section{DISCUSSION AND CONCLUSION}

 In order to compare the influence of pressure on the superconductivity and short-range magnetic order in
Pr$_{1.2}$La$_{0.7}$Ce$_{0.1}$CuO$_{4}$, the pressure dependences of the short-range magnetic order temperature  $T_{\rm so}$ ($T_{\rm so}^{onset}$) and the SC transition temperature $T_{\rm c}$ are shown in Fig. 10a. Moreover, the pressure dependences of $\chi^{''}$ and $\chi_{\rm ZFC}$, normalised to its ambient pressure value, are shown in Fig. 10b. The most essential findings of the present work are the following: 1) The short-range magnetic order temperature $T_{\rm so}$ in Pr$_{1.2}$La$_{0.7}$Ce$_{0.1}$CuO$_{4}$ exhibits the strong and non-linear positive pressure effect. $T_{\rm so}$ is nearly constant in the pressure range between 0 - 0.8 GPa, increasing at higher pressure. For the highest applied pressure of $p$ ${\simeq}$ 2.3 GPa the $T_{\rm so}$ increases by ${\sim}$ 15 K. To the best of our knowledge this is the largest pressure effect observed in electron-doped cuprates and the first report on the hydrostatic pressure effect on magnetism. Note that the magnetic order remains short range even at $p$ ${\simeq}$ 2.3 GPa. 
2) The SC transition temperature $T_{\rm c}$ Pr$_{1.2}$La$_{0.7}$Ce$_{0.1}$CuO$_{4}$ shows the non-linear negative pressure effect. Likewise $T_{\rm so}$, $T_{\rm c}$  is constant in the pressure range between 0 - 0.8 GPa. However, above $p$ ${\simeq}$ 0.8 GPa, $T_{\rm c}$ decreases instead of the observed increase for $T_{\rm so}$. For the applied pressure of $p$ ${\simeq}$ 2.1 GPa, $T_{\rm c}$ decreases by ${\sim}$ 3-4 K. Moreover, the diamagnetic response decreases substantially and non-linearly with increasing pressure and above $p$ ${\simeq}$ 2.1 GPa, no diamagnetic signal is observed using bulk sensitive techniques. This means that the superconductivity in Pr$_{1.2}$La$_{0.7}$Ce$_{0.1}$CuO$_{4}$ is either suppressed fully or becomes filamentary above $p$ ${\simeq}$ 2.1 GPa. Antagonistic pressure behaviour between $T_{\rm so}$ and $T_{\rm c}$, gives direct evidence for the competition between superconductivity and the short-range magnetic order in Pr$_{1.2}$La$_{0.7}$Ce$_{0.1}$CuO$_{4}$. It is interesting that the magnetic fraction is nearly pressure independent, while the SC fraction reduces substantially. This is different from hole-doped cuprates with the static spin and charge order, in which ${\mu}$SR observes phase separation between magnetism and superconductivity \cite{Guguchia1,Guguchia2016}. In addition, the magnetic order in Pr$_{1.2}$La$_{0.7}$Ce$_{0.1}$CuO$_{4}$ remains short range up to the highest applied pressure, which probably suggests that the superconductivity is not suppressed above $p$ ${\simeq}$ 2.1 GPa, but it becomes filamentary and still inhibits the long-range magnetic order in the system.

  Additional results provide important clues of how pressure may induce the strong changes in magnetic and SC properties. In the closely related compounds Ln$_{2-x}$Ce$_{x}$CuO$_{4}$ ($Ln$ = Nd, Sm, and Eu) (where isovalent substitution of Nd by Sm and Eu simulates a condition generally referred to as ''chemical pressure'', that is, both in-plane and out-of-plane lattice constants become small), ARPES and first-principles electronic-structure calculations revealed that the chemical pressure (the variation of the in-plane and out-of-plane lattice constants) has a great influence on the electronic structures and on the Fermi-surface shape \cite{Ikeda}. Namely, reduction of the in-plane lattice constant causes the reduction of -$t^{'}$/$t$, where $t$ and $t^{'}$ denote transfer integrals between the nearest-neighbor and next-nearest-neighbor Cu sites and enhancement of AFM order. It was suggested that reduced -$t^{'}$/$t$ enhances the AFM order through strengthened nesting, resulting in the increased gap in the nodal region. As we have shown through neutron powder diffraction experiments, hydrostatic pressure causes the continuous reduction of the in-plane and out-of-plane lattice constants in T$^{'}$-Pr$_{1.2}$La$_{0.7}$Ce$_{0.1}$CuO$_{4}$, which, according to the above discussion, might also improve nesting and enhance the AFM order, leading to the increase of $T_{\rm so}$ and decrease of $T_{\rm c}$. Thus, we suggest that the short-range magnetic order and the strong increase of $T_{\rm so}$ is intrinsic and is controlled by the Fermi surface properties of Pr$_{1.2}$La$_{0.7}$Ce$_{0.1}$CuO$_{4}$.

\begin{figure}[t!]
\centering
\includegraphics[width=1.0\linewidth]{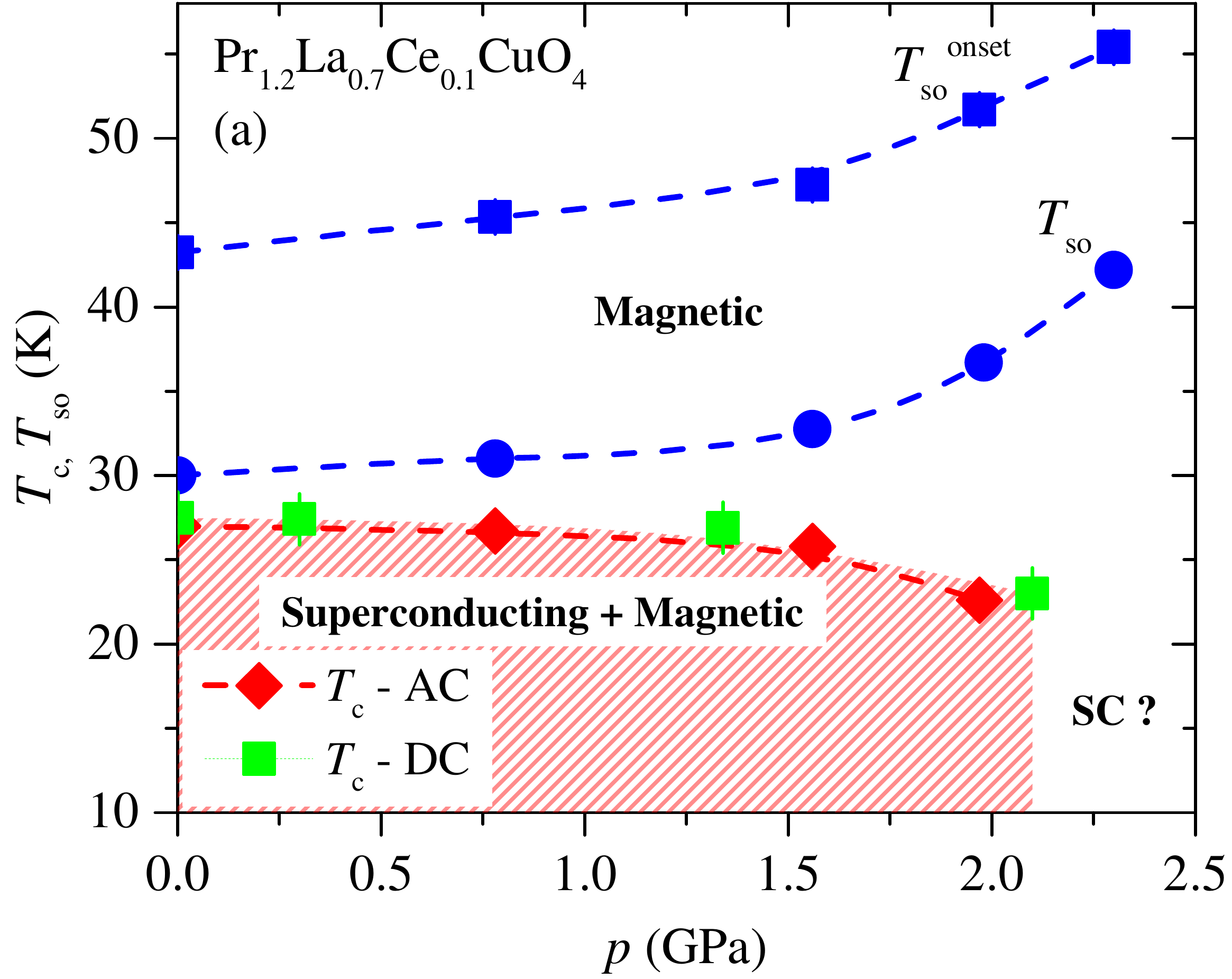}
\includegraphics[width=1.1\linewidth]{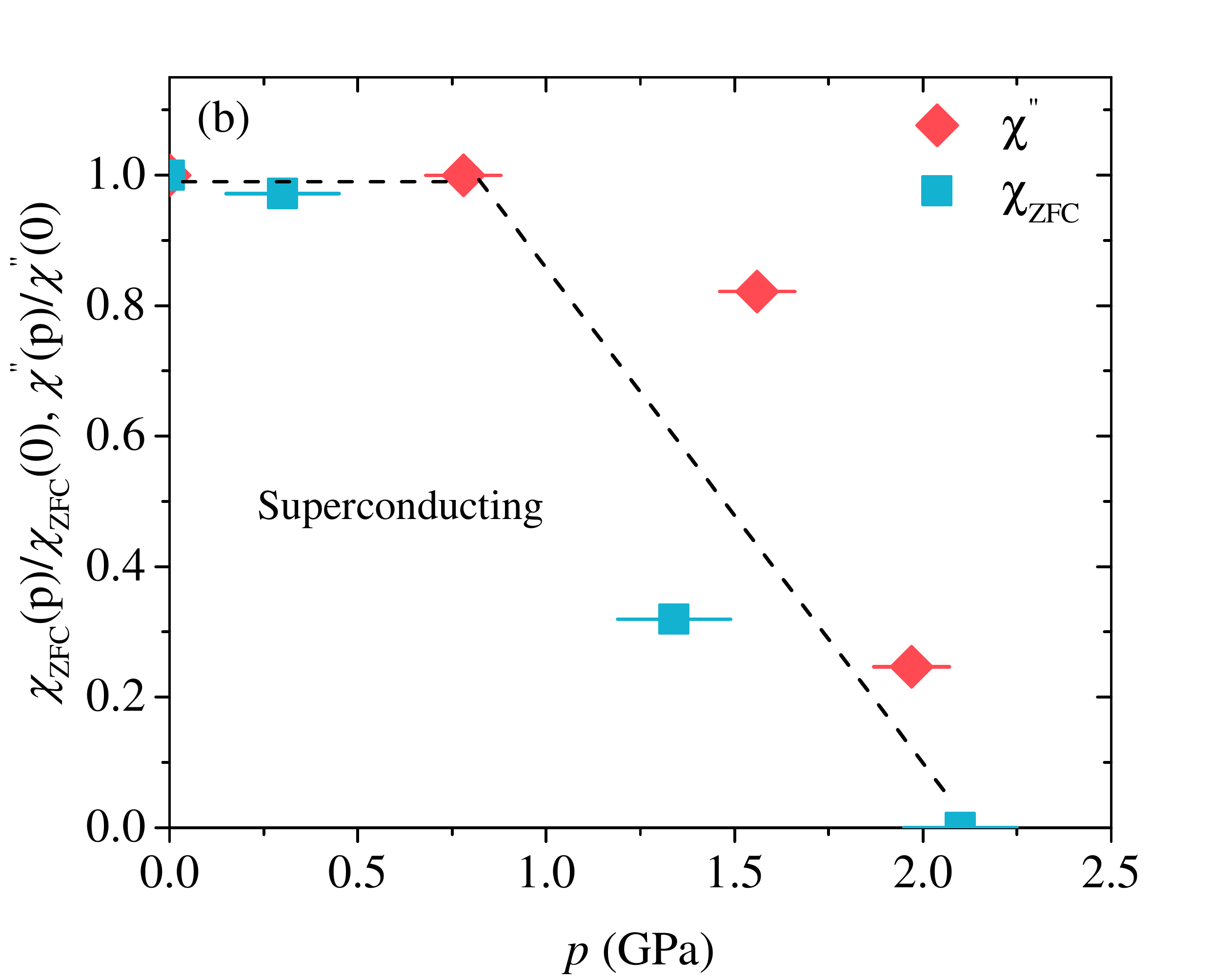}
\vspace{-0.6cm}
\caption{ (Color online) (a) The superconducting transition temperature $T_{\rm c}$ and
the magnetic ordering temperatures $T_{\rm so}$ and $T_{\rm so}^{onset}$ of Pr$_{1.2}$La$_{0.7}$Ce$_{0.1}$CuO$_{4}$, obtained from DC/AC susceptibilities and ${\mu}$SR experiments, are plotted as a function of pressure. (b) The pressure dependence of $\chi^{''}$ and $\chi_{\rm ZFC}$, normalised to its ambient pressure value. The dashed lines are guides to the eye.}
\label{fig1}
\end{figure}

 It is important to draw parallel between the observation of short-range magnetic order at $T_{\rm so}$ ${\simeq}$ 45 K in Pr$_{1.2}$La$_{0.7}$Ce$_{0.1}$CuO$_{4}$ and the recent findings in the related SC $n$-type cuprate Nd$_{2-x}$Ce$_{x}$CuO$_{4}$ \cite{SilvaNeto,Astuto,Hinton,Lee2014,Ishii}. The existence of charge ordering with a short correlation length (15 to 27 \r{A}) was reported for Nd$_{2-x}$Ce$_{x}$CuO$_{4}$ \cite{SilvaNeto}. Time-resolved reflectivity studies in SC Nd$_{2-x}$Ce$_{x}$CuO$_{4}$ show the presence of a fluctuating order below ${\sim}$ 75 K, although they could not determine which electronic degrees of freedom (i.e., charge or spin) were responsible for such order \cite{Hinton}. In this SC Nd$_{2-x}$Ce$_{x}$CuO$_{4}$, resonant inelastic x-ray scattering measurements \cite{Lee2014,Ishii} have also recently shown the presence of an inelastic mode. All these observations suggest that some new phases are emerging together with superconductivity in electron-doped cuprates. The current results are in line with this suggestion. Additionally, we demonstrated the competing nature of the short-range magnetically ordered state to superconductivity in Pr$_{1.2}$La$_{0.7}$Ce$_{0.1}$CuO$_{4}$. Our discovery of short-range magnetic order may contribute in understanding the complex electron-doped cuprate phase diagram.

   In conclusion, hydrostatic pressure effects on short-range magnetic order and superconductivity in electron doped cuprate T$^{'}$-Pr$_{1.2}$La$_{0.7}$Ce$_{0.1}$CuO$_{4}$ were investigated by combining high pressure ${\mu}$SR and AC as well as DC susceptibility experiments. At all applied pressures, nearly the whole sample volume exhibits the short-range magnetic order. The order temperature $T_{\rm so}$ exhibits a large positive pressure effect which has never been observed before for electron doped cuprates.  Furthermore, the observed pressure induced shifts of $T_{\rm so}$ and the superconducting transition temperature $T_{\rm c}$ have opposite signs. 
Moreover, the strong reduction of the lattice constants $a$ and $c$ is observed under pressure. However, no indication of the pressure induced phase transition from T$^{'}$ to T structure is observed up to the maximum applied pressure of $p$ = 11 GPa. These experiments establish the short range magnetic order as an intrinsic part of the phase diagram and as a new competing phase in electron-doped cuprate superconductor T$^{'}$-Pr$_{1.2}$La$_{0.7}$Ce$_{0.1}$CuO$_{4}$. The observed pressure effects may be interpreted by assuming a strong pressure induced changes on the electronic structure and the Fermi surface through the variation of the lattice constants. Namely, the improved nesting upon pressure might be a possible explanation for the enhancement of $T_{\rm so}$ and suppression of $T_{\rm c}$.

 \textbf{\section{Acknowledgments}}

The ${\mu}$SR experiments were carried out at the Swiss Muon Source (S${\mu}$S) Paul Scherrer Insitute, Villigen, Switzerland. A portion of this research used resources at the Spallation Neutron Source, a DOE Office of Science User Facility operated by the Oak Ridge National Laboratory. Z.G. gratefully acknowledges the financial support by the Swiss National Science Foundation (SNFfellowship P2ZHP2-161980). Work at Department of Physics of Columbia University is supported by US NSF DMR-1436095 (DMREF) and NSF DMR-1610633. Neutron powder diffraction data collection and analysis in the Billinge-group were supported by  U.S. Department of Energy, Office of Science, Office of Basic Energy Sciences (DOE-BES) under contract No. DE-SC00112704. Z.G. thanks H. Keller, A. Shengelaya, H. Luetkens, A.N. Pasupathy, and R.M. Fernandes for useful discussions. J.C. gratefully acknowledges the financial support by the Swiss National Science Foundation. The authors T.A. and Y.K. acknowledges support from the JSPS KAKENHI Grant Number 23540399 and by MEXT KAKENHI Grant Number 23108004. 

\newpage 


\end{document}